\documentclass[doublecol]{epl2}

\usepackage{epsfig,graphicx,times}

\usepackage{amssymb}
\usepackage{amsmath}
\usepackage{graphicx}
\usepackage{dcolumn}
\usepackage{bm}

\title{Radiation- and Phonon-Bottleneck-Induced Tunneling in the Fe$_8$ Single-Molecule Magnet}

\author{ M. Bal\inst{1} \and Jonathan R. Friedman\inst{1}\thanks{E-mail:\email{ jrfriedman@amherst.edu}} \and W. Chen\inst{2} \and M. T. Tuominen\inst{3} \and C. C. Beedle\inst{4} \and E. M. Rumberger\inst{4} \and D. N. Hendrickson\inst{4}}
\shortauthor{M. Bal \etal}
\institute{
\inst{1} Department of Physics, Amherst College, Amherst, MA 01002-5000 \\
\inst{2}Department of Physics and Astronomy, Stony Brook University, Stony Brook, NY 11794\\
\inst{3}Department of Physics, University of Massachusetts, Amherst, MA 01003\\
\inst{4}Department of Chemistry and Biochemistry, University of California at San Diego, La Jolla, CA 92093
}

\abstract{
\noindent We measure  magnetization changes in  a single crystal
of the single-molecule magnet Fe$_8$ when exposed to intense, short ($\le$20 $\mu$s) pulses of microwave radiation resonant with the $m$ = 10 to 9 transition.  We find that radiation induces a phonon bottleneck in the system with a time scale of $\sim$5 $\mu$s.  The phonon bottleneck, in turn, drives the spin dynamics, allowing observation of thermally assisted resonant tunneling between spin states at the 100-ns time scale.  Detailed numerical simulations quantitatively reproduce the data and yield a spin-phonon relaxation time of $T_1\sim$40 ns.
}
\begin{document}
\maketitle

%\pacs{75.45.+j, 75.50.Xx, 76.30.-v}

%\narrowtext
Since the discovery of resonant tunneling between spin states more than a decade ago \cite{33, 497}, single-molecule magnets (SMMs) have been intensively studied.  In the past few years, much effort has focused on the behavior of SMMs  in the presence of microwave/millimeter-wave radiation as a way to understand the fundamental spin dynamics and the coupling of the spin to its environment \cite{283, 339, 563, 562, 584, 585, 602, 581, 583, bahr}.  This work has been motivated, in part, by the possibility that these systems could serve as qubits \cite{298}.

At low temperatures ($\lesssim$10 K), the Fe$_8$ SMM behaves as a spin-10 object with uniaxial anisotropy.  The spin dynamics can be well described by the Hamiltonian

\begin{equation}
{\cal H} =  - DS_z^2  + E(S_x^2-S_y^2)+C(S_+^4+S_-^4)- g\mu _B
\vec{S} \cdot  \vec{H}, \label{Ham}
\end{equation}

\noindent where the anisotropy constants $D$, $E$, and $C$ are
0.292 K, 0.046 K, and $-2.9$ $\times$ 10$^{-5}$K, respectively,
and $g$ = 2 \cite{162, 188, 184}. The first term produces a double-well potential for the spin's orientation, making the ``up'' and ``down'' directions (relative to the z axis) lowest in energy and producing a  $\sim25$ K barrier between the two orientations \cite{91}.  A magnetic field $H$ along the z axis makes one well lower in energy than the other, as illustrated in the insets to Fig.\ref{sims}.  There are 2$S$+1 = 21 energy levels for this $S=$10 system.  The second and third terms
in Eq.~\ref{Ham} break its cylindrical rotational symmetry and result in
tunneling between levels. Resonant tunneling occurs when the magnetic field causes levels in opposite wells to align.

Recent studies of the magnetization dynamics in a radiation field have attempted to observe radiation-induced dynamics (such as Rabi oscillations) that would allow a direct determination of the lifetimes of excited spin states and the dephasing time $T_2$.  %, a parameter that is crucial for quantum computing applications.
In previous work on Fe$_8$, we showed that such efforts are complicated by the fact that resonant radiation heats the sample and drives the spins and lattice out of equilibrium on the millisecond timescale \cite{562}, a process that can be quantitatively described \cite{585}.  To circumvent this heating phenomenon, we have done experiments at a much shorter time scale in which an Fe$_8$ sample is  subjected to intense, short microwave pulses.  Such a time-domain technique allows us to investigate the magnetization dynamics as a temporal sequence of transitions between spin levels.  We observe the development of a phonon bottleneck that limits the spin-phonon relaxation and plays an essential role in the magnetization dynamics.  The phonon bottleneck drives subsequent relaxation and allows us to directly observe the thermally assisted resonant tunneling process at time scales of $\lesssim$100 ns.  Through detailed numerical simulations, we deduce an excited-state lifetime ($T_1$) of $\sim$40 ns for this system.

A single crystal of Fe$_8$ was mounted in a cylindrical resonant cavity with Q$\sim$6200.  The radiation field $H_1$ at the sample position was estimated to be 1 Oe.  The sample was fixed above a lithographically defined inductive pick-up loop, with the crystal's b axis parallel to the plane of the loop to maximize flux coupling.  The emf induced in the loop when the sample's magnetization changed was measured with a SQUID voltmeter coupled to a room-temperature amplifier.  Further details of the experimental setup are described elsewhere \cite{584}.  To characterize our sample, we performed standard electron-spin resonance (ESR) reflection spectroscopy (not shown), which allowed us to determine the sample's orientation angles $\theta$ = 37.5$^\circ$ and $\phi$ = 138$^\circ$ (defined by $\vec{S} \cdot  \vec{H}=
H(S_z \cos \theta  + S_x \sin \theta \cos \varphi  + S_y \sin \theta \sin \varphi )$), consistent with the directly measured orientation of the crystal.

\begin{figure}[tbh]
\centering
\includegraphics[width=80mm]{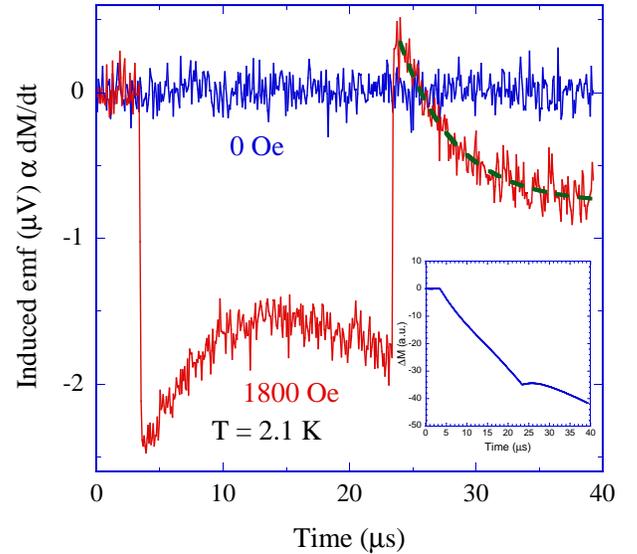} %\vskip 20 pt
\caption{(Color online) Emf  ($\propto dM/dt$) induced in our pick-up loop as a function of time when a 20-$\mu$s radiation pulse of frequency 117.566 GHz is applied to an Fe$_8$ crystal at 2.1 K in zero field and in a field of 1800 Oe, as designated.  The dashed line shows an exponential fit to the 1800-Oe data after the end of the radiation pulse.  The inset shows the change in magnetization $\Delta$M obtained from numerically integrating the 1800-Oe data.
\label{longpulse} }
\end{figure}

Figure \ref{longpulse} shows the induced emf produced by our sample when a 20-$\mu$s  pulse of 117.566-GHz radiation is applied.  At a magnetic field of 1800 Oe, where the radiation is resonant with the transition between states $m$ = 10 and $m$ = 9, we observe a clear signal (proportional to dM/dt), punctuated by sudden jumps at the times when the radiation is turned on and off.  In contrast, data taken at zero field, where the radiation does not couple to the sample, shows no detectable signal.  For reference, the inset of Fig.~\ref{longpulse} shows the 1800-Oe data after numerical integration to obtain $\Delta$M as a function of time.  The data indicate that during the radiation pulse the magnetization decreases, as population is pumped out of the $m$ = 10 state.  After the radiation is turned off, the magnetization continues a downward trend, an effect we have previously characterized as being due to the spins and lattice having been driven out of equilibrium by the radiation \cite{585, 562}.  On top of this overall downward trend, there is a decay immediately after the radiation is turned on or off.  We fit the data after the radiation is turned off to an exponential decay plus a constant term (the latter to account for the slow heating that occurs with a time scale of $\sim$1 ms \cite{585, 562}).  The results of the fit are shown by the dashed line in the figure and yield a time constant of 4.4(3) $\mu$s.  We typically find a decay time of $\sim$5 $\mu$s for this sample, with no systematic dependence on temperature, field or radiation power.  We interpret this relaxation as the signature of a phonon bottleneck \cite{86, 603} in which emission of phonons during the decay from $m$ = 9 to 10 leads to a build up in the  population of phonons resonant with that transition.  The timescale for spins to populate the excited state during the radiation pulse or return to the ground state after the pulse is determined by the time for the phonon distribution to build up or decay away, respectively, the phonon-bottleneck lifetime $\tau_{pb}$.  Evidence for phonon-bottleneck effects have been seen in other SMMs \cite{586} and the effect has been suggested to occur in Fe$_8$ \cite{581}.  Ours is the first direct measurement of $\tau_{pb}$ in a SMM system.  $\tau_{pb}$ represents the time scale either for the resonant phonons to thermalize via nonlinear processes within the crystal or to escape into the environment of the cryostat.

\begin{figure}[htb]
\centering
\includegraphics[width=80mm]{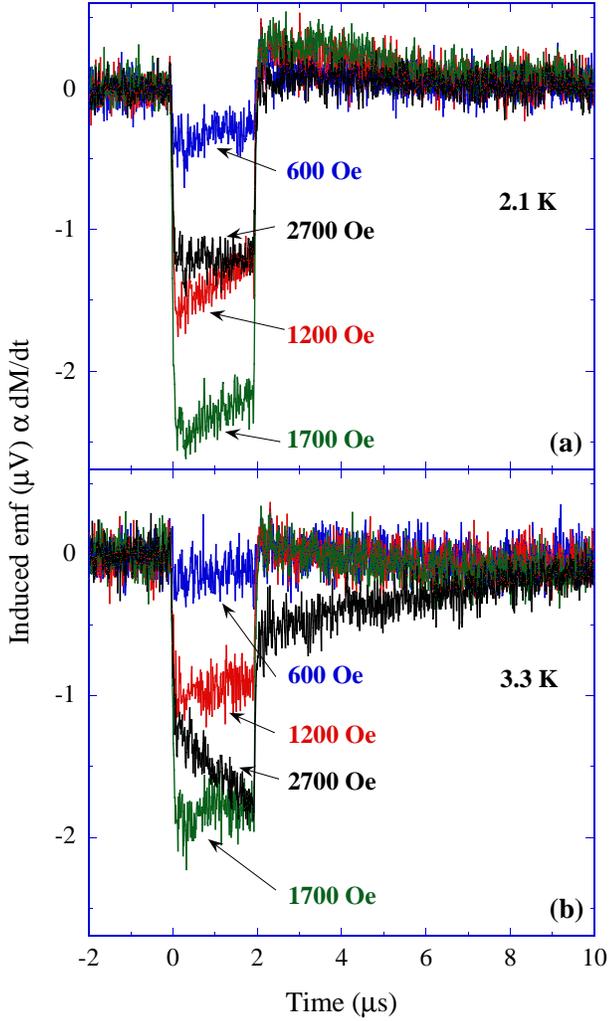} %\vskip 20 pt
\caption{(Color online) Induced emf for 2-$\mu$s radiation pulses applied to the Fe$_8$ sample at a) 2.1 K and b) 3.3 K at several values of magnetic field, as noted.  We define t=0 to coincide with the time when the radiation is turned on.  Data at many values of magnetic field were taken; only a small subset is shown for clarity of presentation.  Each curve represents the average of 12800 oscilloscope traces in a) and 16000 traces in b).  The time between pulses was 20 ms.
\label{shortpulse} }
\end{figure}

\begin{figure}[htb]
\centering
\includegraphics[width=80mm]{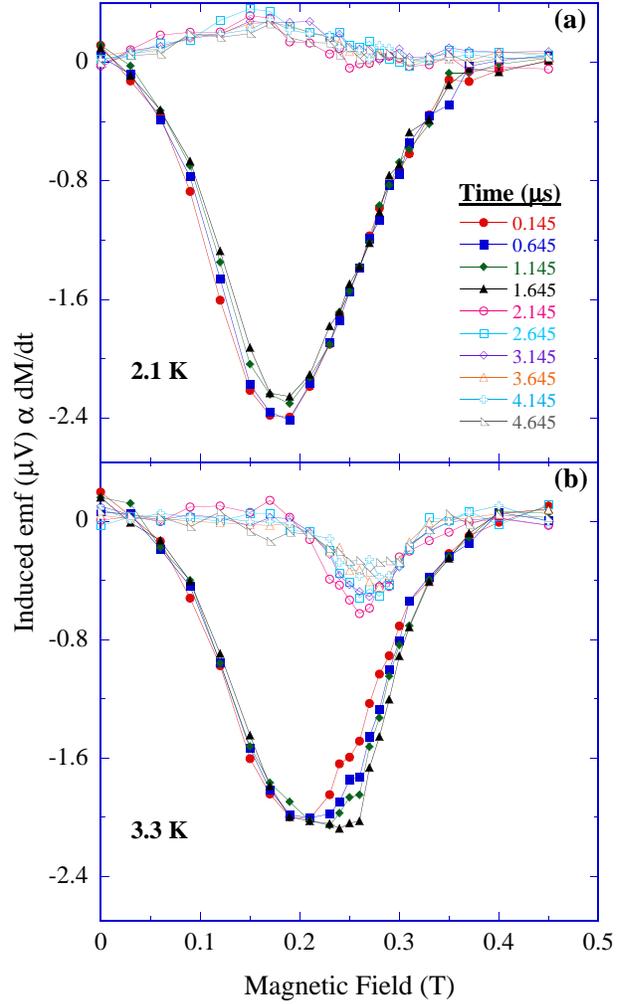} %\vskip 20 pt
\caption{(Color online) Induced emf as a function of magnetic field for several points in time after the onset of a 2-$\mu$s radiation pulse, as noted, obtained from the data in Fig.~\ref{shortpulse} and similar data not shown, at a) 2.1 K and b) 3.3 K, averaged over a 100-ns time window.
\label{lineshape} }
\end{figure}

\begin{figure}[htb]
\includegraphics[width=80mm]{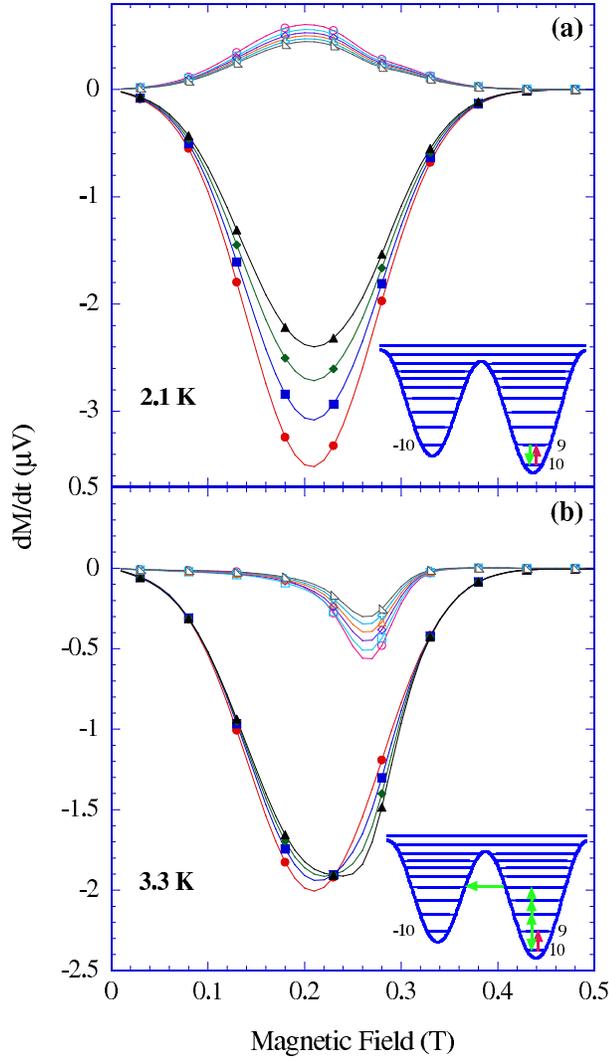}
\caption{(Color online) Results of simulations.  The calculated values of $dM/dt$ as a function of field at various times after the onset of a radiation pulse.  Symbols are the same as in Fig.~\ref{lineshape}.  The values of $dM/dt$ have been scaled by experimentally determined parameters to have the units of induced emf ($\mu$V).  The insets show schematically the dominant level transitions determined by the simulations.  Red (green) arrows represent primarily photon- (phonon-) driven transitions.
\label{sims} }
\end{figure}

In order to study spin dynamics before the complication of lattice heating sets in, we focus on shorter pulses of 2-$\mu$s duration.  Some examples of the dM/dt signal as a function of time for different values of magnetic field at 2.1 K (3.3 K) are shown in Fig.~\ref{shortpulse}a (\ref{shortpulse}b).  To elucidate the dynamics, we analyze the data by taking several time slices of the curves in Fig.~\ref{shortpulse} (as well as similar curves at other field values, not shown) to measure the dM/dt ``lineshape'' as a function of field.  The results of such an analysis are shown in Fig.~\ref{lineshape}.  Figure \ref{lineshape}a shows the data at 2.1 K.  At this temperature, the dynamics are essentially restricted to the $m$ = 10 and 9 levels (see Fig.~\ref{sims}a inset).  When the radiation is on (solid symbols), dM/dt is negative: the magnetization decreases as population is moved from $m$ = 10 to 9.  After the radiation is turned off (open symbols), dM/dt becomes positive as the population returns to the ground state at the $\sim$5-$\mu$s phonon bottleneck time scale.

When we raise the temperature to 3.3 K, the dynamics become richer, as shown in Fig.~\ref{lineshape}b.  Immediately after the radiation is turned on (0.145 $\mu$s), the lineshape looks approximately symmetric, similar to the 2.1 K data.  However, at later times a shoulder develops on the high-field side of the curve.  After the radiation is turned off, $dM/dt$ in the vicinity of the shoulder remains negative with a minimum at $\sim$2650 Oe.  Thus, the magnetization continues to decrease after the radiation is turned off.  The position of this minimum corresponds to the field at which resonant tunneling occurs between levels in opposite wells for Fe$_8$ at the crystal's orientation.

These results can be understood in terms of a thermally assisted resonant tunneling process driven by the radiation/phonon bottleneck, as follows (Fig.~\ref{sims} insets).  The radiation promotes some population to the $m$ = 9 state, where it stays for some time because of the phonon bottleneck.  If the temperature is low (e.g.~2.1 K), there are insufficient thermal phonons to significantly populate any other level on the order of $\tau_{pb}$ and the system decays back to equilibrium.  At the higher temperature (e.g.~3.3 K), thermal phonons are able to excite the system to higher levels (m = 8, 7, \ldots).  Some population is then transferred into the opposite well by tunneling.  Even a small population change between wells can lead to a relatively large signal since $\Delta$m is large ($\sim$10--20) for tunneling transitions.  The role of the phonon bottleneck is essential to the dynamics:  it slows down the transitions between the $m$ = 9 and 10 states, which is primarily responsible for the 2.1-K results, while at higher temperatures (e.g. 3.3 K) thermally assisted tunneling effects can be observed on faster ($\sim$100-ns) timescales because there is no bottleneck for transitions between higher-lying states.
%\begin{figure}[htb]
%\centering
%\includegraphics[width=80mm]{Fig3.eps} %\vskip 20 pt
%\caption{(Color online) Induced emf as a function of magnetic field for several points in time after the onset of a 2$\mu$s radiation pulse, as noted, obtained from the data in Fig.~\ref{shortpulse} and similar data not shown, at a) 2.1 K and b) 3.3 K, averaged over a 100-ns time window.
%\label{lineshape} }
%\end{figure}


We can quantitatively model these dynamics through a master-equation approach \cite{24,221,113,168} that includes radiative transitions, phonon transitions and treats the population of phonons resonant with the 10-to-9 transition as a dynamical variable.  We work in the spin's energy eigenbasis, which incorporates tunneling effects automatically through the fact that the eigenstates of Eq.~\ref{Ham} are superpositions of $m$ levels.  We neglect off-diagonal elements in the density matrix, which is a good approximation except extremely close to a tunneling resonance (when two levels are within a natural linewidth of each other).  The rate of change of the population of the \emph{ith} energy eigenstate is given by:

\begin{equation}
\frac{{dP_i }}{{dt}} =  \sum\limits_{\scriptstyle j = 1 \hfill \atop
  \scriptstyle i \ne j \hfill}^{21} {-(\gamma _{ij}^{+}  + \gamma _{ij}^{- }  + w_{ij} )} P_i  + {(\gamma _{ji}^{+ }  + \gamma _{ji}^{- }  + w_{ji} )} P_j. \label{master}
\end{equation}
The first term represents transitions from the \emph{ith} level to the others and the second term represents transitions to the \emph{ith} level.  The populations are normalized so that $\sum\limits_iP_i=1$. The photon ($w_{ij}$) and phonon ($\gamma _{ij}^{\pm }$) transition rates are calculated using golden-rule expressions \cite{86, 221}:

\begin{equation}
\begin{aligned}
w_{ij}  = &\frac{{\pi(H_1 g\mu _B )^2 }}{{2\hbar ^2 }}\left| {\left\langle i \right|S_x \left| j \right\rangle}\right|^2  F(\omega) \\
\gamma _{ij}^{\pm }  = &\frac{{D^2 }}{{24\pi \rho c_s^5 \hbar ^4 }}\left| {\left\langle i \right|\left\{ {S_ \pm  ,S_z } \right\}\left| j \right\rangle } \right|^2 \left( {\varepsilon _i  - \varepsilon _j } \right)^3 N_{ij}^{ph}, \label{rates}
\end{aligned}
\end{equation}
where $S_\pm$ are the standard spin raising and lowering operators, $\varepsilon_i$ is the energy of level $\left| i \right\rangle$, $\rho$ is the mass density, and $c_s$ is the transverse speed of sound.  $F(\omega)$ is the resonance lineshape function, which we take to be a Gaussian of width $\sigma$.  We do not consider collective spin-phonon interactions \cite{590}.
The rates $w_{ij}$ are appreciable only during the radiation pulse \cite{note2}.  For the phonon transition rates in Eq.~\ref{rates}, $N_{ij}^{ph}=n_{ij}$ (+1) for absorption (emission) of phonons, where $n_{ij}$ is the number of phonons (per spin) resonant with the $i\rightarrow j$ transition.  For most of the phonon transitions we set $N_{ij}^{ph}$ to its thermal equilibrium value $(e^{(\varepsilon _i  - \varepsilon _j)/k_B T}-1)^{-1}$. For transitions between the $m$ = 10  ($\left| g \right\rangle$) and $m$ = 9 ($\left| e \right\rangle$) states, we treat the phonon number as a variable $n_{pb}$, the phonon-bottleneck number:

\begin{equation}
\begin{aligned}
 \frac{{dn_{pb} }}{{dt}} =  &- P_{g} \frac{{\left( {\gamma _{g,e}^{+ }  + \gamma _{g,e}^{- } } \right)}}{{N_{g,e}^{ph} }}n_{pb}  \\
&+ P_{e} \frac{{\left( {\gamma _{e,g}^{+ }  + \gamma _{e,g}^{- } } \right)}}{{N_{e,g}^{ph} }}\left( {n_{pb}  + 1} \right) - \frac{{\left( {n_{pb}  - N_{e,g}^{ph} } \right)}}{{\tau}_{pb} } \\
\label{pb}
\end{aligned}
\end{equation}

Here the first term on the right-hand side is the rate of phonon absorption, the second term, the rate of phonon emission, including spontaneous emission, and the last term, the rate at which the phonons decay toward their equilibrium population.  We numerically solve Eqs.~\ref{master} and \ref{pb} for the populations $P_i$ and $n_{pb}$ as functions of time, with thermal equilibrium values used as initial conditions. The rate of magnetization change is calculated using $\frac{dM}{dt} = \sum\limits_{i = 1}^{21} {\left\langle i \right|\overrightarrow S  \cdot \frac{\overrightarrow H }{\left| {\overrightarrow H } \right|}\left| i \right\rangle } \frac{{dP_i }}{{dt}}$, where $\overrightarrow S  \cdot \frac{\overrightarrow H } {\left| {\overrightarrow H } \right|}$ is the projection of the spin operator along the external-field (measurement) direction.  Finally, the results are convoluted with a Gaussian of width 200 Oe to account for inhomogeneous broadening of the resonant tunneling features.  The results of these simulations are shown in Fig.~\ref{sims}.

In performing the simulations, we used the known Hamiltonian and the spectroscopically determined orientation of the sample.  The position, width and depth of  the main (photon/bottleneck) peak are respectively determined by parameters $D$, $\sigma$ and $H_1$; these we set to 0.290 K, 650 Oe and 1.0 Oe, respectively, each in good agreement with its independently measured value ($D$ and $\sigma$ determined from ESR spectra.).  We set the phonon bottleneck time to the measured value of $\tau_{pb} = 5\,\mu$s (Fig.~\ref{longpulse}).  We treat the temperature $T$ as a constant on the timescale probed here.  We set $\rho$ = 1920 kg/m$^3$ \cite{588}\.  The time dependence of the magnetization in our simulations is extremely sensitive to the value of $c_s$ (because of the factor $c_s^5$ in Eq.~\ref{rates}).  We found $c_s$ = 670 m/s to give the best agreement between the simulations and our 3.3-K data.  This value is 16\% smaller than the value extracted from specific heat measurements \cite{589}; the discrepancy could arise from the fact that specific heat measures the average speed of sound, not the transverse component.  It has been proposed that spin-phonon transitions in which $\Delta m =$ 2 could provide an additional relaxation mechanism and increase the relaxation rate \cite{168, 113}.  We were unable to adequately simulate our data using such a model.  Using the above parameters, we calculated the system's relaxation rate in the absence of radiation near zero magnetic field at several temperatures and obtained results consistent with those found in ac susceptibility measurements on the same material \cite{193}. From Eq.~\ref{rates} and the fit value of $c_s$, we deduce a zero-temperature (i.e. spontaneous emission) lifetime of the $m$ = 9 excited state of $T_1\sim$ 40 ns. The simulations also allow us to deduce the dominant level transitions, shown in the insets to Fig.~\ref{sims}, with the dominant tunneling transition occurring between levels $m$ = 6 and -7.

Our results are in contrast to some recent work by another group studying the same material. Petukhov et al.~\cite{581} have suggested that a phonon bottleneck is responsible for the $>$1-ms radiation-induced dynamics in Fe$_8$.  The time scale of those observations, however, corresponds to the heating effect characterized in \cite {562,585}.  Very recently, Bahr \emph{et al.} \cite{bahr} used a pump-probe technique to study the magnetization dynamics of Fe$_8$ exposed to pulsed radiation.  They inferred from their data that $T_1\sim 2 \mu$s for the $m$ = 9 state, nearly two orders of magnitude slower than the value determined in this study, but close to our value of $\tau_{pb}$.  We found it impossible to simulate the tunneling results (Fig.~\ref{lineshape}b) with such a long $T_1$, essentially because the thermal phonon transition rates must be slower than 1/$T_1$, preventing thermally assisted tunneling dynamics from occurring on the timescale of $\sim$100 ns when $T_1\sim 2 \mu$s, even if $\Delta m =$ 2 processes are included.  The discrepancy between the two sets of experiments is the subject of ongoing investigation.

%\begin{figure}[tbh]
%\centering
%\includegraphics[width=80mm]{Fig4.eps} %\vskip 20 pt
%\caption{(Color online) Results of simulations.  The calculated values of $dM/dt$ as a function of field as various times after the onset of a radiation pulse.  Symbols are the same as in Fig.~\ref{lineshape}.  The values of $dM/dt$ have been scaled by experimentally determined parameters to have the units of induced emf ($\mu$V).  The insets show schematically the dominant level transitions determined by the simulations.  Red (green) arrows represent primarily photon- (phonon-) driven transitions.
%\label{sims} }
%\end{figure}

Our simulations capture all of the qualitative features of our data and give very good quantitative agreement with the 3.3-K results and approximate agreement with the 2.1-K data.  We note that the peak in the 2.1-K data is shifted slightly toward lower field relative to its position at 3.3 K, an effect that has been ascribed to  dipole fields \cite{438}.  While the simulations reasonably reproduce the 2.1-K data at fields below $\sim$1000 Oe, they show a larger signal and stronger time dependence near the minimum than the actual data.  We attribute this to the fact that when the sample is on resonance with the cavity (i.e. near the peak), its absorption effectively reduces the cavity Q, lowering the value of $H_1$ and therefore the  photon absorption rate.  This effect is more pronounced at lower temperature when the ground-state population is larger.  It will also affect the time dependence as the populations change with photon absorption.

The line width $\sigma$ of the photon-absorption resonances represents the effect of inhomogeneous broadening.  Our observation of tunneling constrains the physical nature of this broadening.  The 3.3-K results require a coincidence between two resonances:  the photon resonance where the photon energy matches that of the 10-to-9 transition and the tunneling resonance in which levels in opposite wells align.  This obviates the possibility that the broadening $\sigma$ could arise primarily from some quasi-static inhomogeneous dipole or hyperfine field since such a random field would shift both resonances equally and could not result in their overlap.  On the other hand, our results are  consistent with a model in which there is a distribution in values of the anisotropy parameter $D$ \cite{438}, wherein only a fraction of the molecules have a value of $D$ that satisfies the coincidence conditions.

In summary, we have measured the fast magnetization dynamics of the single-molecule magnet Fe$_8$ subject to short pulses of intense microwave radiation.  We observe the development of a phonon bottleneck with a decay time of $\sim$5 $\mu$s.  The bottleneck can drive resonant tunneling between excited states when thermal phonons are available to populate these states, i.e. at sufficiently high temperature.  At low temperatures, the dynamics are restricted to the the two states resonant with the radiation and dominated by the phonon-bottleneck time scale.  Our detailed numerical simulations are in good agreement with the data and imply an excited state lifetime of $\sim$40 ns.

\acknowledgments
We thank E. Chudnovsky, D. Garanin, D. Hall, L. Hunter, M. Sarachik and W. Wernsdorfer for useful conversations.
We also thank D. Krause, A. Anderson and G. Gallo for their technical contributions to this
study and E. H. da Silva Neto, N. Schmandt, J. Rasowsky and J. Atkinson for their help in acquiring and analyzing some of
the data. Support for this work was provided by the National
Science Foundation under grant nos.~CCF-0218469 and DMR-0449516, and by the Amherst College Dean of Faculty.


\begin{thebibliography}{10}
\bibitem{note2} The shape of the radiation pulse used in the simulations includes the cavity's measured rise/ring-down time of $t_{cav}=9 $ ns and is defined by

\begin{equation}
\Pi \left( t \right) = \left\{ \begin{aligned}
 &0 \, &t < 0  \\
 &1 - e^{ - \frac{t}{t_{cav} }} \, &0  \le t \le t_{pulse}\\
 &\left( 1 - e^{\frac{- t_{pulse}}{t_{cav}}} \right) e^{- \frac{\left( t - t_{pulse}\right)}{t_{cav} }} \,&t > t_{pulse}   \\
 \end{aligned} \right.
 \nonumber
 \end{equation}
\end{thebibliography}
\end{document}